\documentclass[aps,prper,twocolumn,floats,a4paper]{revtex4-1}
\usepackage[T1]{fontenc}
\usepackage[utf8]{inputenc}
\usepackage{graphicx}
\usepackage{color}
\usepackage{enumitem}
\usepackage[cmex10]{amsmath}

\usepackage{amsmath}
\usepackage{amssymb}

\begin{document}

\title{Spreading height and critical conditions for the collapse of
  turbulent fountains in stratified media}

\author{L. G. Sarasua, D. Freire, C. Cabeza, Arturo C. Marti}
\affiliation{Instituto  de F\'\i sica,  
             Facultad de Ciencias,
Universidad de la Rep\'ublica, Igu\'a 4225, Montevideo, Uruguay}

\begin{abstract}
Axisymmetric fountains in stratified environments rise until reaching
a maximum height, where the vertical momentum vanishes, and then falls
and spread radially as an annular plume following a well-known top-hat
profile.  Here, firstly, we generalize the model of Morton et
al. (Proc. R. Soc. Lond. A \textbf{234}, 1, 1956), in order to
correctly determine the dependence of the maximum height and the
spreading height with the parameters involved. We obtain the critical
conditions for the collapse of the fountain, \textit{i.e.} when the
jet falls up to the source level, and show that the spreading height
must be expressed as a function of at least two parameters.  To
improve the quantitative agreement with the experiments we modify the
criterion to take the mixing process in the down flow into
account. Numerical simulations were implemented to estimate the
parameter values that characterizes this merging. We show that our
generalized model agrees very well with the experimental measurements.
\end{abstract}

\maketitle


\section{Introduction}
 A fountain is a vertical buoyant jet in which the buoyancy force and
 the jet initial velocity act in opposite directions.  If the buoyancy
 force acts in the same direction of the jet velocity it is said the
 flow is a plume.  Fountains and plumes are encountered frequently in
 nature and technical applications such as selective withdrawal,
 desalination plants, and the replenishment of magma chambers.  Since
 the dynamical behavior of fluids within stratified media presents a
 problem of considerable interest across a number of fields, turbulent
 fountains and plumes in uniform and stratified mediums have been the
 subject of investigation for decades \cite{Woods, Kaye, Burridge,
   Turner1969, Turner,Bloomfield1998, Bloomfield1999,Lin, Hunt2001,
   Hunt2005,Richards,Camassa,Carroll,Ezhova}.

The dynamics of the fountain in a stratified medium can be schematized
as follows. At an initial stage the fountain decelerates due to both
the entrainment of ambient fluid and the negative buoyancy force, and
it reaches a maximum height where the momentum is zero. Then the flow
reverses direction and falls as an annular plume around the fountain
core. Depending on the initial fluxes of momentum and buoyancy and the
stratification profile, the fountain spreads outwards at a non zero
spreading height above the source level or the flow totally collapses,
i.e.  falls to the level of the source.

The theoretical description of turbulent fountains in the quasi-steady
regime can be done on the basis of the influential work of Morton,
Taylor and Turner \cite{Morton1956, Morton1959, Morton1973}, who
derived equations (so called {\it MTT} equations) for the evolution of
volume, momentum, and buoyancy fluxes in fountains. In deriving these
equations, it is assumed that the horizontal velocity at which the
ambient fluid enters into the fountain is proportional to the vertical
velocity in the fountain, with a proportionality constant $\alpha$
called entrainment coefficient.  Although successful in predicting the
evolution in a uniform ambient or the maximum height in plumes
\cite{Kaye}, the $MTT$ equations do not describe the dynamics after
the vertical velocity reverses its direction.

 Bloomfield and Kerr proposed that the spreading height, $z_s$, can be
 obtained matching to the height where the fluid of the environment
 has the density of the fluid at the maximum height, $z_m$, and used
 this condition to obtain estimations of $z_m$ and $z_s$ combining
 different models \cite{Bloomfield1998}. This may be considered as a
 first estimation because the mentioned condition does not take into
 account the mixing between the jet and ambient fluids in the downflow
 that occurs after the fountain reverses direction. In a later work
 Bloomfield and Kerr developed a theoretical model to predict the
 maximal spreading height \cite{Bloomfield2000} based on the equations
 derived by McDougall for an axisymmetric fountain in an homogeneous
 fluid \cite{McDougall}. In this model the authors assumed entrainment
 equations which depends on three entrainment constants.

A while later, Kaminski et al. \cite{Kaminski}, developed an
expression for the entrainment parameter $\alpha$ depending on three
parameters that can be determined using the experimental data.  A
comparison between the predictions based on this expression and the
experimental data were given in \cite{Carazzo} for the case of
homogeneous mediums.  Mehaddi et. al. \cite{Mehaddi} conducted a study
of fountains in stratified environments and obtained expression for
the maximum height, although the spreading behavior of the fountain
was not considered in this investigation. Papanicolau et al. conducted
an experimental study on the collapse and spreading of turbulent
fountains and performed a comparison with those obtained in
\cite{Bloomfield1998}.  As it has been pointed out by various authors
\cite{Turner1966,Telford,Reeuwijk}, to assume a constant $\alpha$ is
an approximation because the entrainment coefficient depends on the
turbulence intensity and as a consequence it can vary along the rise
of the fountain.

In this work we study models for the rise and spreading of a fountain
in a stratified medium. Although new equations are introduced in the
models, a constant value of $\alpha$ is assumed. Our aim is to develop
a model which is able to describe the dependence of of $z_m$, $z_s$
with the parameters that determine the flow and the critical
conditions for the total collapse of the fountain. The work is
organized as follows. In Sec. II we give an account of the
\textit{MTT} equations and discuss the capability of the top-hat
version of these equations to determine the spreading level. In Sec.
III we present new equations to avoid a drawback detected in the
Gaussian version of the \textit{MTT} equations. In section IV we
discuss a model which includes the top-hat and Gaussian models as
particular cases and present the numerical results obtained with the
different models. In section V our conclusions are summarized.

\section{The MTT entrainment equations}

We begin considering the equations derived by Morton, Taylor and
Turner \cite{Morton1956, Morton1973} in the Boussinesq approximation
for top-hat axysimmetric steady fountains in a linearly stratified
environment.  In the model the fountain is characterized by an ambient
fluid, initially quiescent, with density $\rho_0 (z)$, being the
density at the bottom $\rho_0(z=0) = \rho_{00}$. A relevant quantity
is the buoyancy frequency $N$, defined as
$N^2=-(g/\rho_{00})(d\rho_0/dz)$.  The vertical jet with radius $b(z)$
it is assumed to enter in that region with velocity $u$. The MTT model
is usually written in terms of vertical flow of mass $W=b^2 u$, the
vertical flow of momentum $M=b^2 u^2$ and bouyancy flux $F= b^2 u g
(\rho_0-\rho)/\rho_{00}$. With these definitions, the model equations
are
\begin{equation}
\frac{dW}{ dz} =2 \alpha M^{1/2}, \ 
\frac{dM^2}{dz}=2 F W,  \
\frac{dF}{dz}= - N^2 W,
\label{mtt}
\end{equation}
where $g$ is the gravitational acceleration. This set of equations
together with the condition according to which $z_s$ is the height
where the fluid of the environment has the density of the fluid at the
maximum height
 \begin{equation}
 \rho (z_m) =\rho_0(z_s),  
 \label{ehs}
 \end{equation}
are sufficient to obtain values of $z_m$ and $z_s$. 
 
 Although the derivation of the \textit{MTT} equations is based on the
 assumptions that the flow is self-similar and the entrainment
 coefficient is constant, this model has been successfully used by
 Bloomfield and Kerr \cite{Bloomfield1998} to predict the maximal
 height $z_m$ of the fountains.  These authors used the condition
 (\ref{ehs}) to obtain an estimation of the spreading heigth $z_s$.
 However, the approach followed in \cite{Bloomfield1998} has the
 inconvenience that it uses two different models to obtain $z_m$ and
 $z_s$. While $z_s$ is estimated by combining Eq.(\ref{ehs}) and the
 integration of \textit{MTT} equations with the initial condition
 $F=0$ at the virtual source, $z_m$ is obtained integrating
 (\ref{mtt}) with the experimental source conditions.

We note that here $z_m$ is the maximal height of the fountain in the
steady regime. It should not be confused with the maximal height
reached in the transitory stage. In this section we analyze the
results that are obtained when the criterion given by Eq.(\ref{ehs})
is used in combination with the integration of (\ref{mtt}) and the
same conditions at the source are imposed to determine $z_m$ and
$z_s$. Thus, these two heights will be determined using the same
model.

\subsection{Dimensionless equations}

We consider now the dimensionless form of the governing equations
(\ref{mtt}) taking the radius of the source $d$ and the velocity at
the source $U$ as the length and velocity scales respectively. Then we
define the dimensionless variables $x=z/d$, $\mathsf{W}=W d^{-2}
U^{-1}$, $\mathsf{M}=M d^{-2} U^{-2}$, $\mathsf{F}=F d^{-1} U^{-3}$,
$\mathsf{N}= d \ U^{-1} N$.  Thus, the dimensionless equations become
\begin{equation}
{d\mathsf{W} \over dx} =2 \alpha \mathsf{M}^{1/2}, \ 
{d\mathsf{M}^2 \over dx}=2 \mathsf{F} \mathsf{W},  \
{d\mathsf{F} \over dx}= -\mathsf{N}^2 \mathsf{W},
\label{mtt2}
\end{equation}
with the initial conditions at the source: 
\begin{equation}
 \mathsf{W}=1, \ \mathsf{M}=1, \ \mathsf{F}= - \textrm{Fr}^{-2},
 \label{ic1}
 \end{equation}
where Fr is the Froude number defined as
Fr$=U/\sqrt{gd(\rho(0)-\rho_{00})/\rho_{00}}$.  As it follows from the
equations (\ref{mtt2}) and initial conditions (\ref{ic1}), the
behavior of the flow is determined by the three independent
dimensionless numbers $\mathcal{N}$, Fr and $\alpha$. This is related
to the fact for the fountain in a linearly stratified medium three
independent characteristic lengths can be defined:
\begin{eqnarray}
l_Q &=&  W_i \alpha^{-1} M_i^{-1/2} \nonumber \\
l_M &=& M_i^{3/4}  \alpha^{-1/2} F_i ^{-1/2} \nonumber \\
l_H &=& F_i^{1/4} \alpha^{-1} N^{-3/4} \label{l3}
\end{eqnarray}
where the subscript \textit{i} refer to the values of $M, W, F$ at the
source \cite{Kaye}. In addition, the radius of the jet at the source
is another relevant longitude. As a consequence, three independent
dimensionless parameters can be constructed by dividing the lengths
$l_Q, l_M, l_H$ by $d$.

In \cite{Bloomfield1998}, \cite{Bloomfield2000} Bloomfield and Kerr argued that the maximal and spreading heights are functions of the form 
\begin{equation}
z_j=f_j(\sigma) M^{-3/4} F^{-1/2}  
\label{hB}
\end{equation}
 where $f_j(\sigma)$ are functions of only $\sigma$, defined as
 $\sigma = M^2 N^2 / F^2$, with $j=m,s$. We notice that this
 expression does not include the dependence on $\alpha$. Owing to the
 fact that $\alpha$ is related to the turbulent mixing, it is a
 function of the Reynolds number $Re=U d /\nu$. Thus, the no
 dependence of equation (\ref{hB}) with $Re$ is a serious
 limitation. In addition, the coefficient $\alpha$ appears in the
 dimensionless parameters (\ref{l3}) that define the flow. As a
 consequence, we expect that $\alpha$ must appear in the expression of
 $z_s$.  In the present work we shall consider the dependence of the
 maximal and the spreading heights on all the relevant parameters that
 determine the flow.

Although the flows are determined by the three independent
dimensionless numbers $\alpha$, $\mathsf{N}$ and Fr, the number of
relevant parameters may be reduced by defining the function
$\mathsf{H}=\mathsf{F}/\alpha$ and $\mathsf{x}=\alpha x$. In this
case, the Eqs.(\ref{mtt2}) become
\begin{equation}
{d\mathsf{W} \over d\mathsf{x}} =2 \mathsf{M}^{1/2}, \ 
{d\mathsf{M}^2 \over d\mathsf{x}}=2 \mathsf{H W},  \
{d\mathsf{H} \over d\mathsf{x}}= -\beta^2 \mathsf{W},
\label{ec2}
\end{equation}
where $\beta=\mathsf{N}/\alpha$. The initial conditions at the source are now
\begin{equation}
\mathsf{W}=1, \ \mathsf{M}=1, \ \mathsf{H}=\ \ \! \eta ^{-1}, \label{ecb2}  
\end{equation}
where $\eta=\alpha$Fr$^{2}$. From the equations (\ref{ec2}), it can be
seen that finally the dimensionless flow only depends on the two
dimensionless parameters $\eta$ and $\beta$. This is a very useful
simplification, because it implies that the fountain regimes can be
represented in a two dimensional diagram.

As already mentioned, the dimensionless spreading height $h_s=z_s/d$
will be obtained as a first approximation using the condition
(\ref{ehs}). With this purpose we write such relation in terms of the
dimensionless variables. From (\ref{ehs}) it follows that
$(\rho(h_m)-\rho_0(h_m))+(\rho_0(h_m)-\rho_{00})=(\rho_0(h_s)-\rho_{00})$
, which can be written as $(\rho(h_m)-\rho_0(h_m))+h_m d\rho_{0}/dz
=h_s d\rho_{0}/dz$. Using the definitions of $F$ and $Q$ we obtain the
value of $h_s$ as a function of the dimensionless maximum height $h_m
= z_m/d$:
\begin{equation}
h_s = h_m + F'(h_m) Q'(h_m)^{-1} N'^{-2}.
\label{xs}
\end{equation}

\subsection{Numerical results}
\begin{figure}
    \includegraphics[width=90mm]{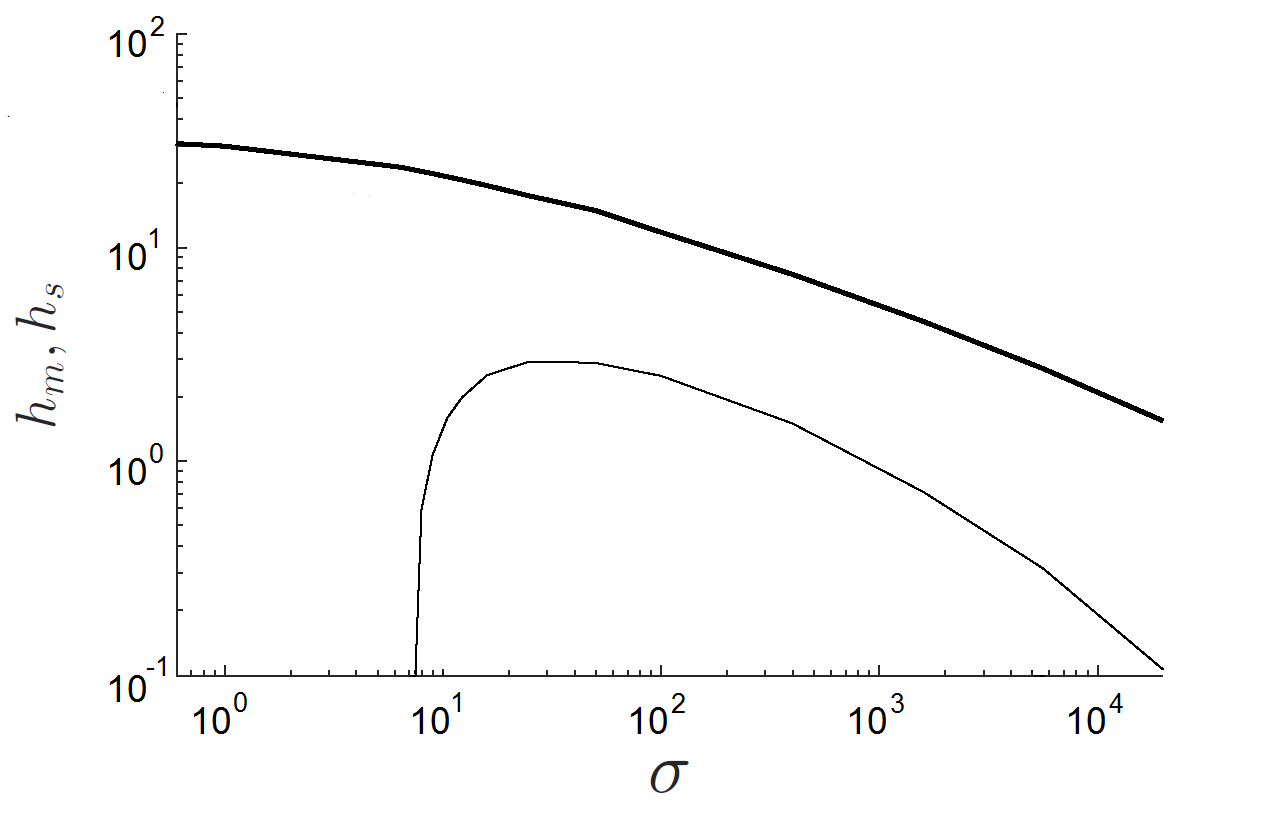}
    \caption{Maximum (thick line) and spreading (thin line)
      dimensionless heights $h_m$, $h_s$ as a function of $\sigma$ for
      $\eta=20$. The collapse of the fountain occurs in this case for
      $\sigma_c=6.89$.}
    \label{fighh}
\end{figure}

\begin{figure}
    \includegraphics[width=90mm]{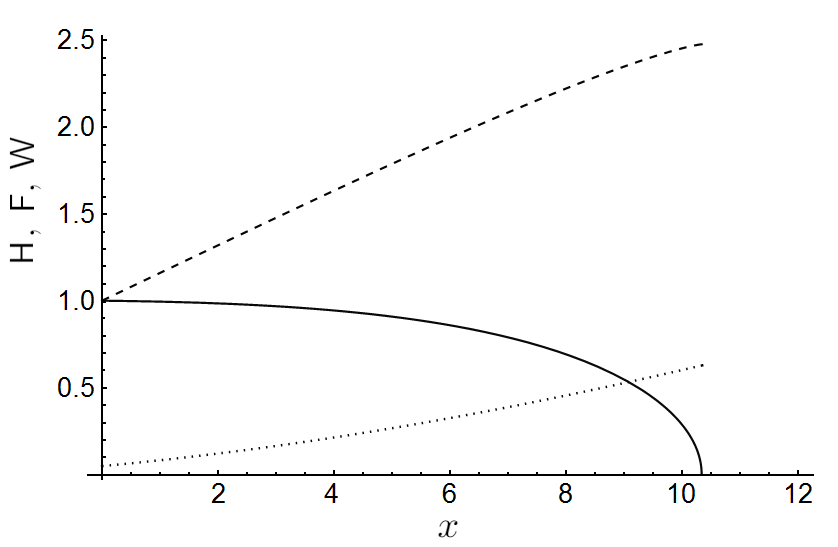}
    \caption{Plot of the functions $\mathsf{M}$ (solid line),
      $\mathsf{W}$ (dashed line) and $-\mathsf{H}$ (dotted line) as a
      function of $x$, for $\mathsf{N}=0.05$, Fr=16.}
    \label{figHFW}
\end{figure}

\begin{figure}
    \includegraphics[width=90mm]{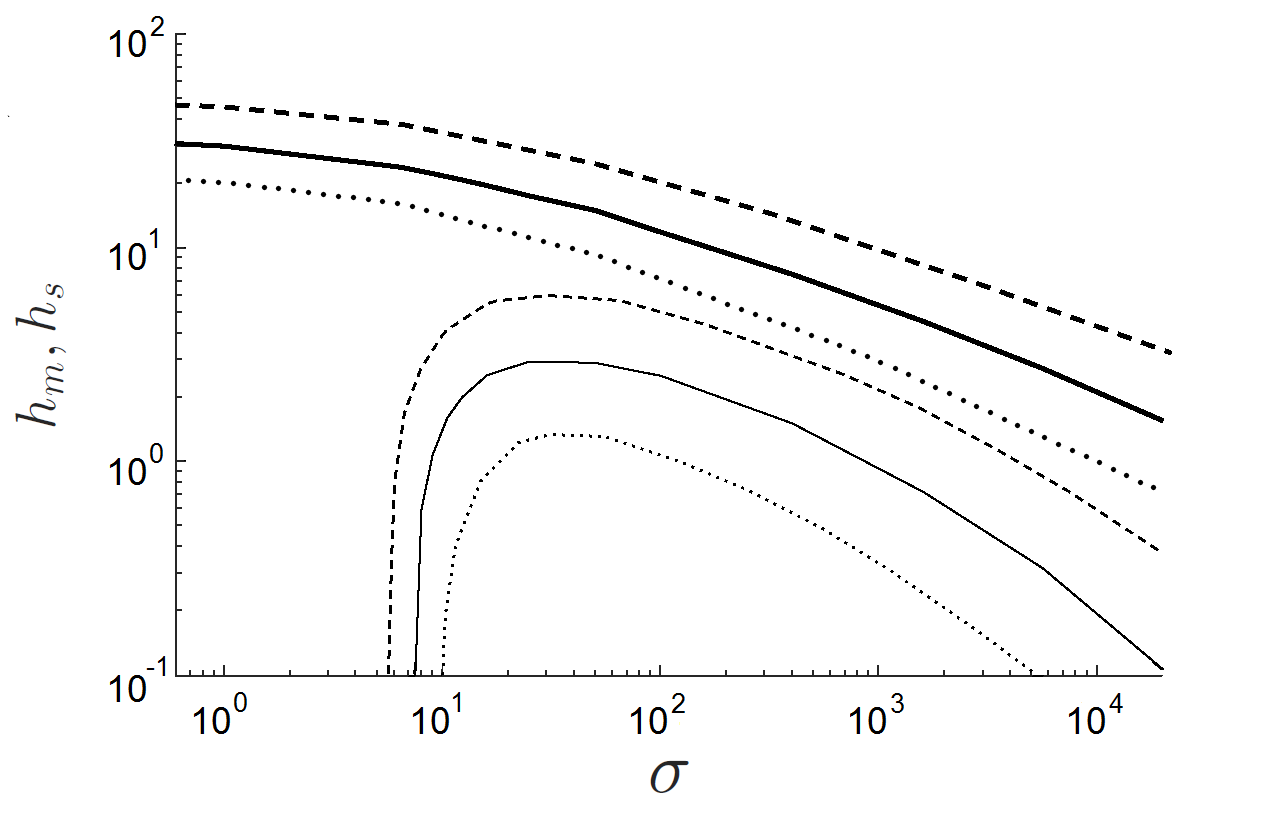}
    \caption{Dimensionless maximal height $h_m$ (thick lines) and
      spreading height $h_s$ (thin lines) as a function of $\sigma$
      for $\eta=40$ (dashed), $20$ (solid) and $10$ (dotted
      line). From the figure the changes of $\sigma_c$ under
      variations of $\eta$ are shown.}
    \label{fighh2}
\end{figure}

In Fig. \ref{fighh} the values of $h_m$ and $h_s$ are shown, which
were obtained integrating the Eqs. (\ref{ec2}), (\ref{xs}) with the
boundary conditions (\ref{ecb2}). It can be seen that the dependence
of $h_m$, $h_s$ with $\sigma$ are in accordance with the experimental
data \cite{Bloomfield1998}. For values of $\sigma$ below a critical
value $\sigma_c$, $h_s$ goes to zero which means that the fountain
collapsed.  We note that the variation of $h_s$ with $\sigma$ is very
abrupt near the collapse at $\sigma = \sigma_c$. This is in agreement
with the experimental data \cite{Bloomfield1998, Papanicolau}. In
Fig. \ref{figHFW} the functions $\mathsf{H, F, W}$ are shown, as a
function of $x$ for the case $\mathsf{N}=0.05$, Fr=16.  The curves of
Fig. \ref{fighh} correspond to $\eta=20$. For other values of $\eta$,
the curves are similar but the value of $\sigma_c$ changes, as it is
plot in Fig. \ref{fighh2}.  These results contradict the hypothesis
(\ref{hB}), according to which $h_s$ is determined by a function that
is proportional to a function that only depends on $\sigma$. Following
this relation, the critical condition $z_s=0$ is given by
$f_j(\sigma)=0$ ( Eq.(\ref{hB})), which would determine a unique value
of $\sigma_c$, independent on Fr, or $\eta$. The results plot in
Fig. \ref{fighh2} show that this is not correct.
 
\begin{figure}
    \includegraphics[width=90mm]{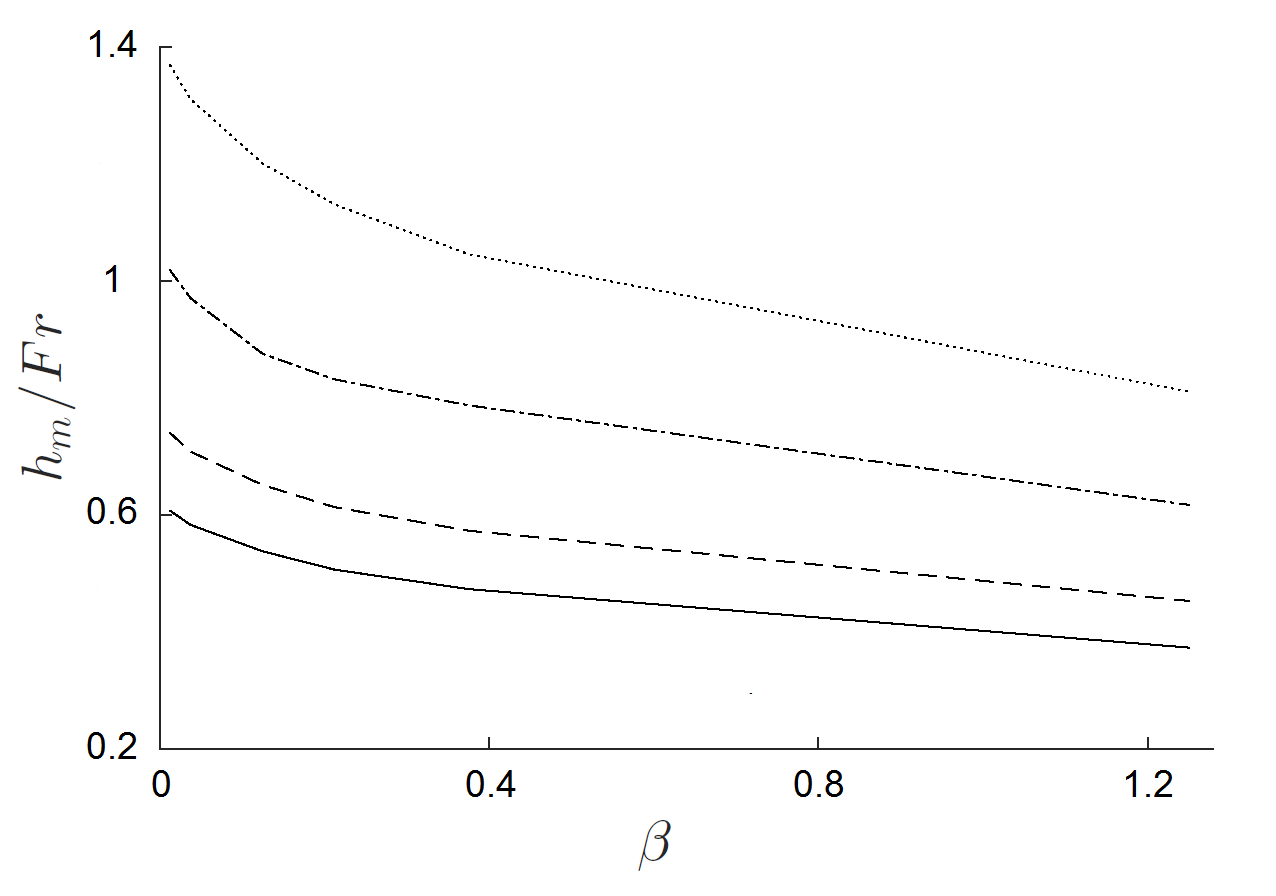}
    \caption{Values of $h_m / Fr$, for $\sigma=25$ (solid line), $100$ (dashed), $400$ (dashdotted) and $900$ (dotted).}
    \label{fignsup}
\end{figure}

We also tested the relation (\ref{hB}) in another way. According to
this relation, $z_m=f_m(\sigma) M^{-3/4} F^{-1/2}$ , which can be
written as $h_m$Fr$^{-1}=f_m(\sigma)$. This means that $h_m$Fr$^{-1}$
should has a unique value for a given value of $\sigma$. In
Fig. \ref{fignsup} curves of $h_m$Fr$^{-1}$ are shown and it can be
seen that they do not overlap.  It could be argued that the
approximations performed to obtain the equations (\ref{ec2}) could
explain the differences with the predictions of
Eq. (\ref{hB}). However, this is not the case. The drawback in
deriving the Eq. (\ref{hB}) is to assume that $h_m$, $h_s$ can be
written in terms of the length scale for an homogeneous environment
\cite{Bloomfield1998} and $\sigma$, without including $\alpha$
explicitly.

\begin{figure}
    \includegraphics[width=90mm]{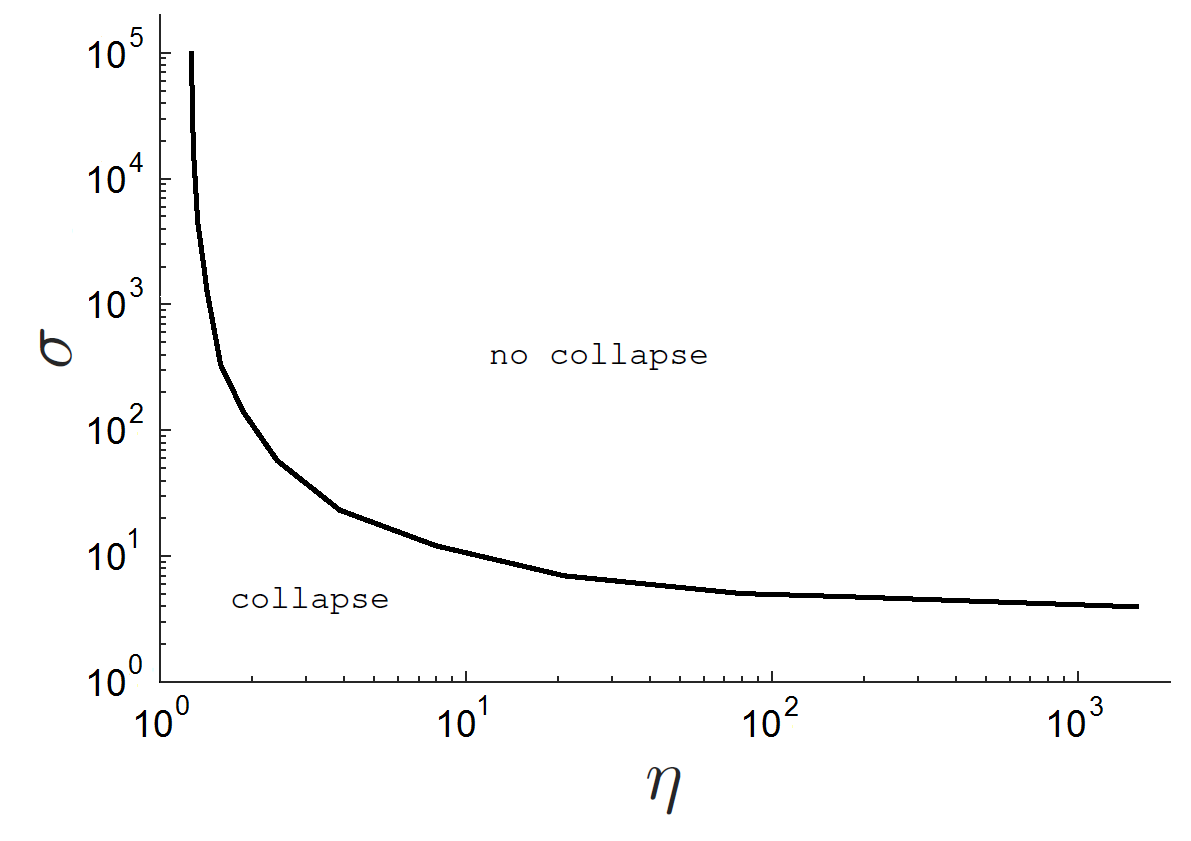}
    \caption{Critical values of $\sigma$ for the collapse of the
      fountain in the top-hat model, using the condition
      (\ref{ehs}). For values of $\sigma$ above the critical line, the
      fountain spreads laterally above the source level.}
    \label{figsigmac}
\end{figure}

In Fig. \ref{figsigmac} the critical value for the collapse of the
fountain as a function of $\eta$ is shown. For this curve it can be
seen that the modification of $\alpha$ can induce a change for the
collapse to the buoyant situation. Since $\sigma = M^2 N^2 / F^2$ is
independent of $\alpha$, an increment of $\eta=\alpha$ Fr$^{2}$ due to
an increment of $\alpha$ maintaining fixed $Fr$ can produce a
transition from the fountain collapse to the fountain spreading above
the source (buoyant regime). Although the typical values of the
entrainment coefficient $\alpha$ are around 0.08 for top-hat profiles
and 0.06 for Gaussian profiles \cite{Bloomfield1998, Burridge}, there
has been a significant variation in the values of $\alpha$ obtained in
distinct experiments \cite{Kaminski, Reeuwijk}. This has been
attributed to differences in the experimental setup and source
conditions \cite{Reeuwijk}. Freire et al. \cite{Freire2010,
  Freire2015} showed that the value of $\alpha$ can be modified
artificially by the introduction of fluctuations. In this work it has
been observed that $z_m$ and thus $\alpha$ can be modified by the use
of meshes at the source, keeping constant the other parameters that
determine the flow. This shows that in certain cases the collapse can
be controlled and avoided by the introduction of fluctuations, which
is an interesting fact from the point of view of practical
applications.

\begin{figure}
    \includegraphics[width=90mm]{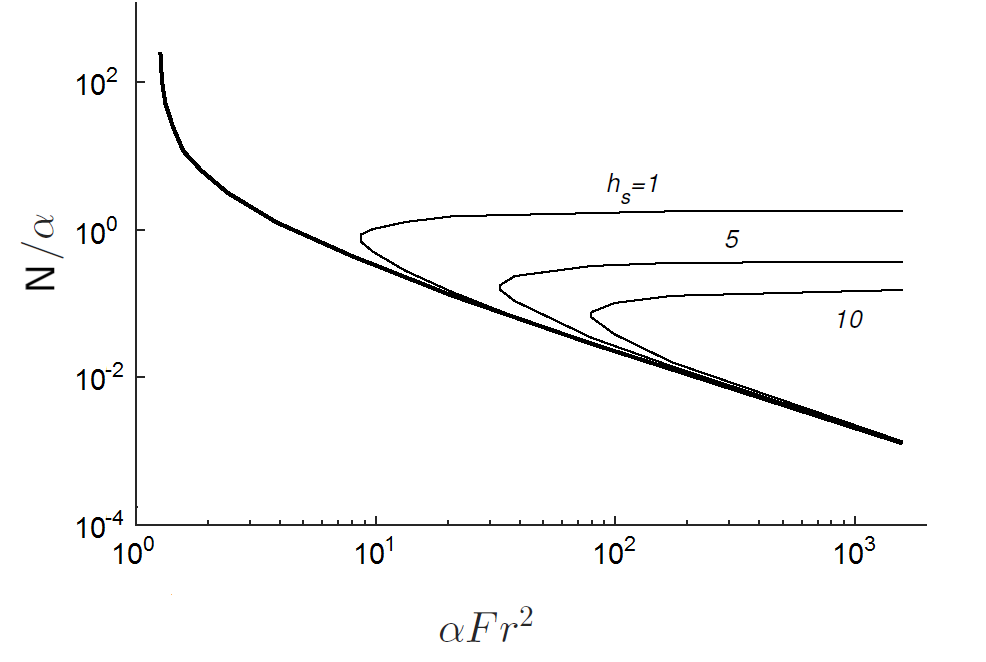}
    \caption{Same as in Fig. \ref{figsigmac} but in function of Fr,
      $\alpha$ and $\mathsf{N}$. Above the critical line (thick line)
      there is no collapse of the fountain. In the figure the curves
      of constant spreading height $h_s$ are also shown, corresponding
      to values indicated on the curves. When the mixing in the
      downflow is considered, these curves are modified.}
    \label{figsigmachs}
\end{figure}

\begin{figure}
    \includegraphics[width=90mm]{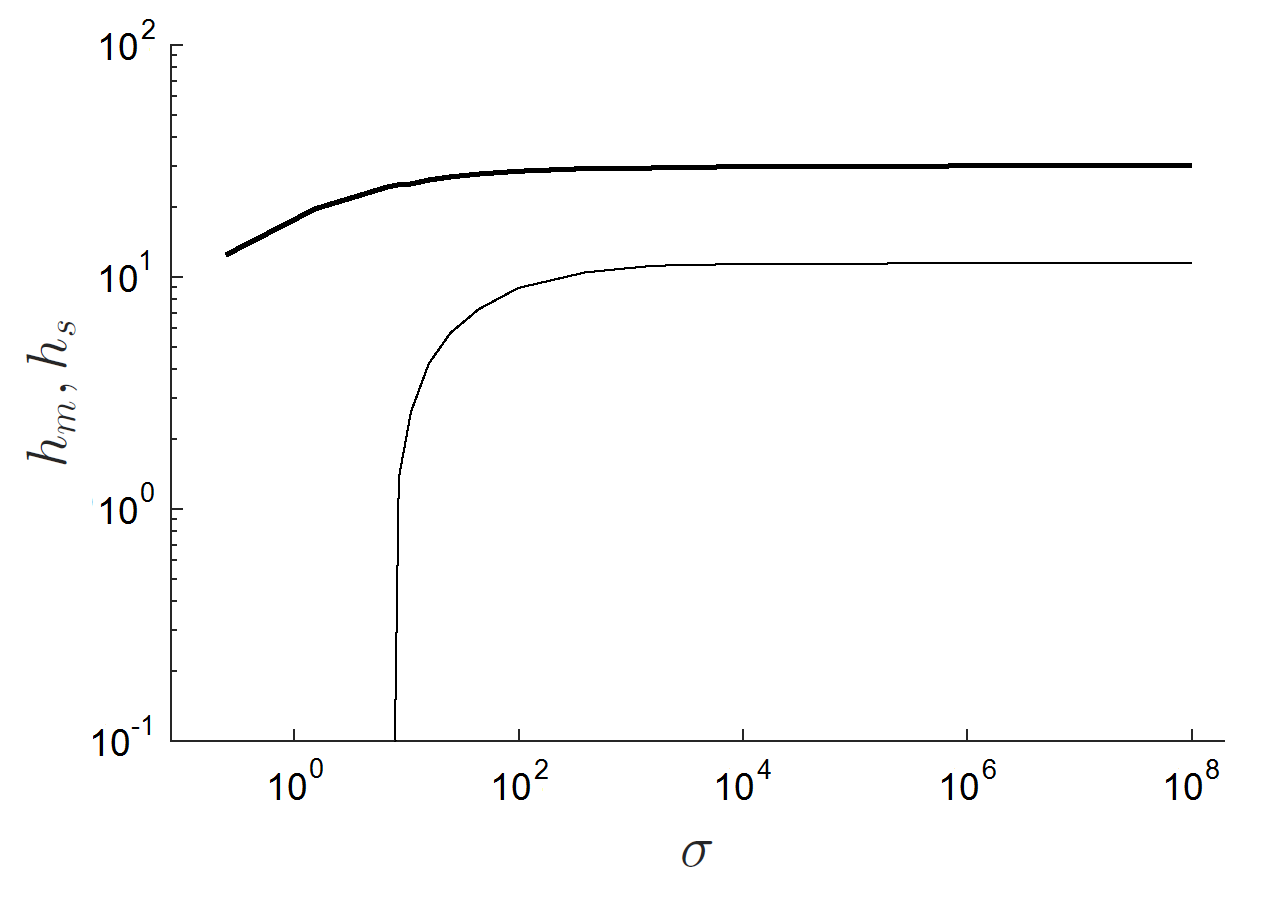}
    \caption{Values of $h_m$ (thick line) and $h_s$ (thin line) as a
      function of $\sigma$ at constant $\beta=0.125$.}
    \label{fighhbeta}
\end{figure}

In the Fig. \ref{figsigmachs} the regions for the collapse or no
collapse of the fountain as a function of $\beta=N'/ \alpha$ and
$\eta=\alpha$Fr$^2$ are shown. In this figure the curves for constant
spreading height $h_s$ are also plotted. In Fig. \ref{fighhbeta} the
values of $h_m$ and $h_s$ as a function of $\sigma$ at a constant
value of $\beta$ are shown.

\begin{figure}
    \includegraphics[width=90mm]{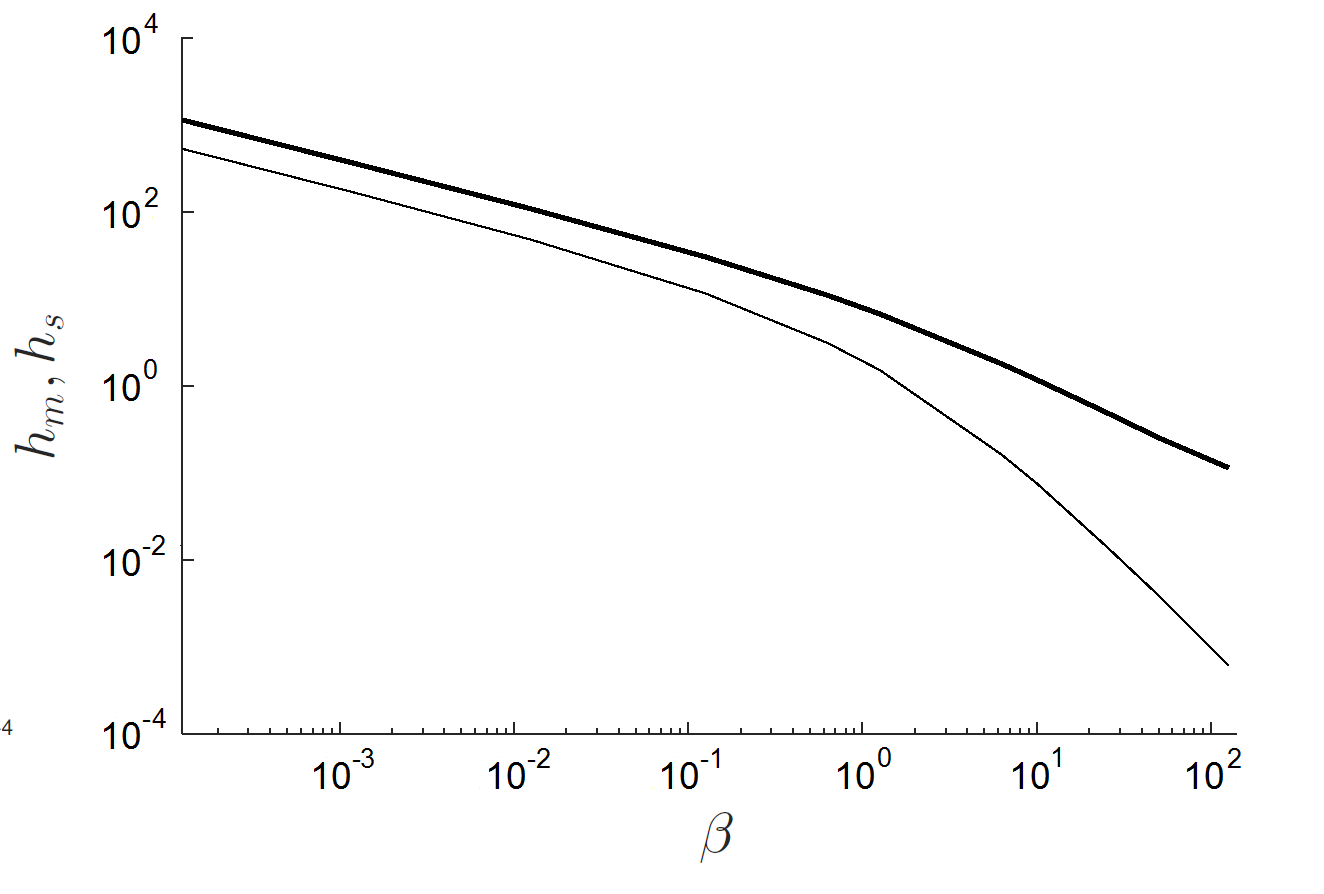}
    \caption{Values of $h_m$ (thick line) and $h_s$ (thin line) as a
      function of $\beta$ for zero buoyancy flux ($\eta \rightarrow
      \infty$). There is no collapse under this condition.}
    \label{fighhbetaf0}
\end{figure}

\subsection{Mixing approximation in the downflow}
\begin{figure}
    \includegraphics[width=90mm]{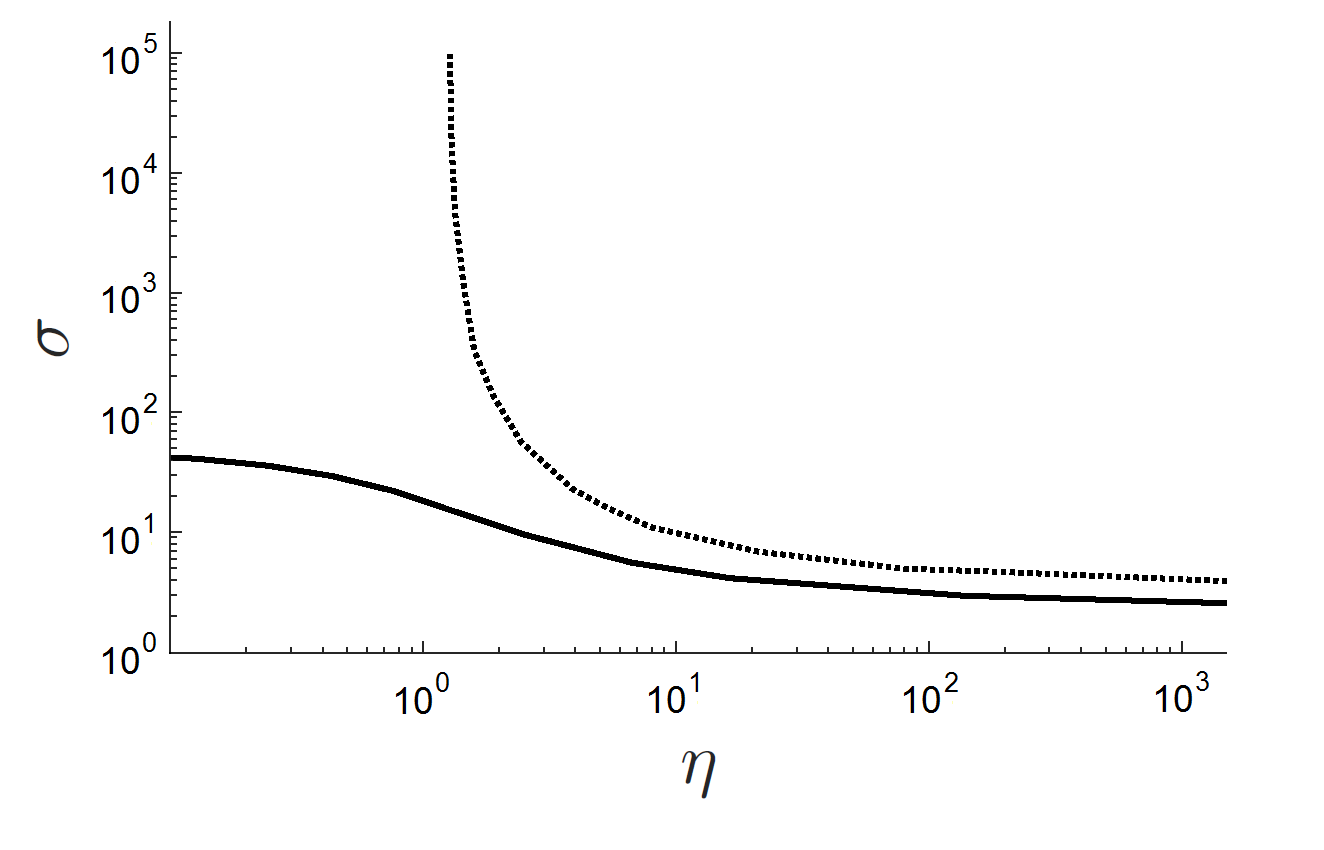}
    \caption{Critical values of $\sigma$ for the collapse of the
      fountain with top-hat profile obtained with the modified
      condition (\ref{nc}) and $\gamma=0.25$ (solid line) to include
      the mixing in the downflow. In the plot it is also included the
      critical line for which $\gamma=0$ (dotted line). }
    \label{figsigmac2}
\end{figure}

As already mentioned, the model considered up to now describes in
correct form the dependence of the fountain flow on the relevant
parameters at a qualitative level, but it does not include the mixing
that takes place after the fountain reversed the velocity
direction. As a consequence, there are important numerical differences
between the critical values predicted by the model and those observed
experimentally. In order to include these second merging effect, we
shall define a parameter $\gamma$ that represents the proportion of
environment fluid to jet fluid that mixes to form the fountain fluid
at the spreading region. As a consequence, the criterion (\ref{ehs})
must be modified in the form:
\begin{equation}
 (1-\gamma) \rho (z_m)+\frac{\gamma}{2}(\rho_0(z_m)+\rho_0(z_s)) =\rho_0(z_s).
\label{ecgamma}
\end{equation}
Here $(\rho_0(z_m)+\rho_0(z_s))/2$ is an average of the environment density in the region around the downflow and $0\leq \gamma<1$. From the above expression we obtain that $\rho_0(h_m)-\rho(h_m)=((1-\gamma/2)/(1-\gamma))(\rho_0(h_m)-\rho_0(h_s))$. Operating in similar form that has been done to derive (\ref{xs}), we obtain that
\begin{equation}
h_s = h_m +2\left(\frac{1-\gamma}{2-\gamma} \right) \mathsf{F}(h_m) \mathsf{Q}(h_m)^{-1} \mathsf{N}^{-2}.
\label{nc}
\end{equation}
In Fig. \ref{figsigmac2} the critical values $\sigma_c$ obtained with
the top-hat model and the condition (\ref{nc}) using $\gamma=0.25$ are
shown. In the figure the values of $\sigma_c$ of for $\gamma=0$ are
also included, which correspond to the original condition
(\ref{ehs}). As it can be seen, the mixture in the downflow reduces
the value of $\sigma_c$, increasing the region for which there is no
collapse of the fountain. This is caused by the fact that the mixing
reduces the density of the fountain, favoring its buoyancy. This
effect is more pronounced for small values of $\eta$.

\section{Gaussian fountains}
In the case of Gaussian fountains, the $MTT$ equations are:
\begin{equation}
{dW \over dz} =2 \alpha M^{1/2}, \ 
{dM^2 \over dz}=4 F W,  \
{dF \over dz}= - 2 N^2 W
\label{mttg}
\end{equation}
\cite{Morton1956}. By integration of these equations and the use of
criteria (\ref{ehs}), values of $h_s$ were obtained for the case of
zero buoyancy flux Fr$^{-1}=0$ ($\eta \rightarrow \infty$), which
means that the density of the fountain equals the environment density
at the source. Under these conditions the obtained values of the
spreading heights were always negative.  Since it is an experimental
evidence that under the condition of zero buoyancy flux the fountains
exhibit always positive values of $h_s$, it follows that the results
given by the model are qualitatively wrong.

\begin{figure}
    \includegraphics[width=90mm]{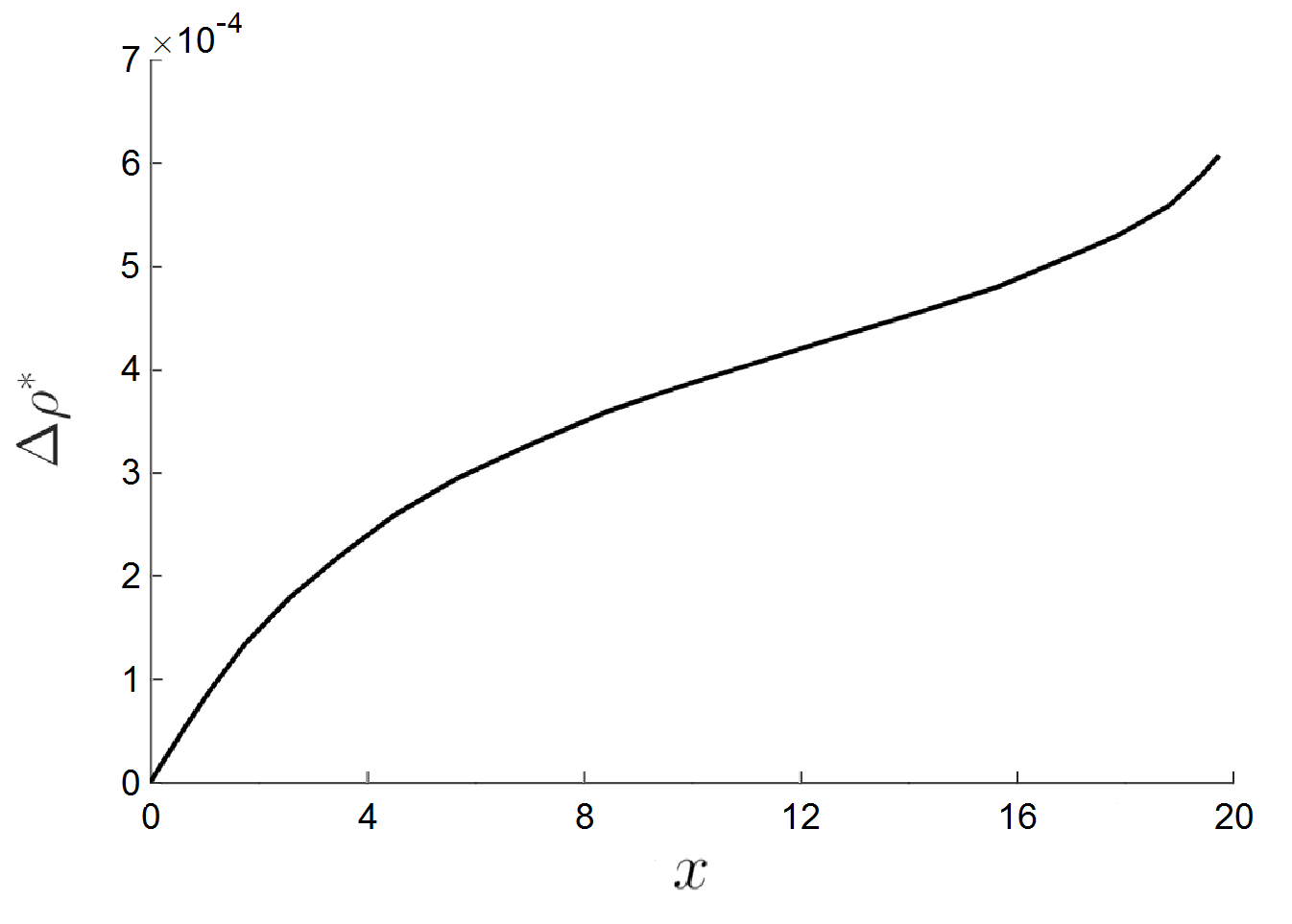}
    \caption{Variation of the jet density $\rho$ with the
      dimensionless height $x$ obtained with de $MTT$ equations for
      the Gaussian profile, with $\mathsf{N}=0.01$, $\alpha=0.06$,
      Fr$^{-1}=0$.}
    \label{figrho}
\end{figure}

In order to visualize the cause of this non expected result we obtain
the value of the fluid density in the jet.  From the definition of
$F$, it follows that
\begin{equation}
\Delta \rho^*=  -\frac{\mathsf{F}}{\mathsf{W}}-\mathsf{N}^2 x    
\end{equation}
where $\Delta \rho^* = (gd/U)(\rho - \rho_{00})/\rho_{00}$ is a
dimensionless measure of the fluid density inside the jet.  In
Fig. \ref{figrho} the variation of $\Delta \rho^*$ with $x$ for
$N=0.01$ and $\alpha=0.06$ is shown. As it can be seen from this
figure, the value of $\rho$ increases with the height in all the
considered cases for Fr$^{-1}=0$, which is an unexpected behavior and
explains why the values of $h_s$ are negative. When the value of
$\rho$ at $h_m$ is larger than $\rho_{00}$, the zero buoyancy level
would be located below the source. When Fr$^{-1} > 0$, the situation
is even worse since the values of $h_s$ are smaller. This shows that
Eqs.(\ref{mttg}) predict an evolution of $\rho$ which is not in
accordance with the experimental observations.

Due to this failure of the model, in the next section we develop a
different equation to describe the changes of $\rho$ caused by the
entrainment process.


\subsection{Density equation}

With the aim to avoid the mentioned drawback of the Gaussian model
described by Eqs. (\ref{mttg}), we developed a new equation to
describe the evolution of the fluid density $\rho$. We assume that in
the entrainment process, a mass $m$ with volume $V$ is mixed with a
mass $\delta m'$ with volume $\delta V'$. Then the final density will
be $\rho_f = (m+\delta m')/(V+\delta V')$. Assuming $\delta m' \ll m$,
$\delta V' \ll V$, then we can write the variation of density as
\begin{equation}
\Delta \rho=\rho_f-\rho_i=(\rho'-\rho_i) \frac{\rho_i \delta m'}{\rho' m}   
\label{ed}
\end{equation}
where $\rho_i=m/V$, $\rho'=\delta m'/\delta V'$. In the case of the
Gaussian jet, the vertical velocity $\tilde{u}$ and the density
$\tilde{\rho}$ are assumed to be functions of the cylindrical
coordinates $(z,r)$ \cite{Morton1956}, with
\begin{equation}
\tilde{u}(z,r) = u(z) e^{-r^2 / b^2} \label{uc}
\end{equation}
and
\begin{equation}
\tilde{\rho}(z,r) = (\rho(z)-\rho_0) e^{-r^2 / b^2}+\rho_0. \label{rc}
\end{equation}
As a consequence of the forms of $\tilde{u}$ and $\tilde{\rho}$ there
is not a precise limit between the core and the outer region of the
jet. In this case, an effective radius can be defined
\cite{Morton1956}.  We shall adopt that the effective radius of the
fountain is $R=b$.
Then considering a vertical cylinder of height $dz$ whose circular
base coincides with the cross section of the jet of radius $b$, the
mass that enters in such cylinder through its base during the time
interval $dt$ is
$$m=\int_0^{b} \tilde{\rho}(z,r) \tilde{u}(r,z) \ 2\pi rdrdt.$$

Substituting in the above expression (\ref{uc}) and (\ref{rc}), and
performing simplification as $\int_0^1 \exp(-2x^2)xdx\approx 0.316$,
we obtain that
\begin{equation}
    m=2\pi u b^2(0.216(\rho-\rho_0)+0.316 \rho_0) dt.
    \label{em}
\end{equation}
On the other hand, the mass that enters through the lateral area is
\cite{Morton1956}
\begin{equation}
\delta m'=2\pi\rho_0 b\alpha u dz dt .
\label{mp}
\end{equation}
The density $\rho_i$ appearing in (\ref{ed}) is
$\rho_i=m/V= V^{-1}\int_0^{b} \tilde{\rho}(r,z) 2\pi r dr dz$, where $V=\pi b^2dz$ is the volume of the cylinder. Substituting (\ref{rc}) in the above expression and performing the already mentioned approximations in the definite integrals we obtain
\begin{equation}
\rho_i=0.632 (\rho-\rho_0)+\rho_0
\label{ri}
\end{equation}

As in the original work of Morton et al. \cite{Morton1956} we consider
that the density variations are small in comparison to $\rho_0$,
$(\rho - \rho_0) \ll \rho_0$. Thus, using this approximation and
substituting the expressions (\ref{em}),(\ref{mp}),(\ref{ri}) in
(\ref{ed}), identifying $\rho'=\rho_0$, $\Delta \rho
=\rho(z+dz)-\rho(z)$, it is obtained that $d\rho=0.835(\rho_0-\rho)
\frac{\alpha}{b}dz$, which can be written as
\begin{equation}
\frac{d\rho}{dz}=0.835(\rho_0-\rho) \alpha \frac{M^{1/2}}{W} 
\label{ed2}
\end{equation}

Defining $G= g (\rho-\rho_0)/\rho_{00}$, Eq.(\ref{ed2}) can be expressed as:
\begin{equation}
\frac{dG}{dz}=-  0.835 \alpha W^{-1}M^{1/2}G+N^2 .
\label{ed3}
\end{equation}
The above equation and the two first equations of (\ref{mtt}) are the
governing equations of our model with Gaussian profile.

We comment that in the case of the top-hat profile, from the equation
(\ref{ed}) the new equation is obtained:
\begin{equation*}
 {dG \over dx}=-2 \alpha {W}^{-1}M^{1/2}G+N^2 .   \label{ecth}
\end{equation*}
This equation was obtained using $\rho_f=\rho(z+dz)$,
$\rho_i=\rho(z)$, $\rho'=\rho_0(z)$, again and operating in similar
way as has been done in section A. In section IV B we consider the
results of a top-hat model including this equation.

\subsection{Non-dimensional equations}
By defining $\mathsf{J}=(d/\alpha U^2)G$, the non-dimensional equations of the Gaussian model can be written as: 
\begin{eqnarray}
{ d\mathsf{W} \over d\mathsf{x}} &=&2 \mathsf{M}^{1/2}, \nonumber \\  
  {d\mathsf{M}^2 \over d\mathsf{x}}&=&-4 \mathsf{J} \mathsf{W}^2,  \nonumber \\
 {d\mathsf{J} \over d\mathsf{x}}=&-& 0.835{\mathsf{W}}^{-1}\mathsf{M^{1/2}J}+\beta^2 
\label{ecg}
\end{eqnarray}
where we used that by definition $F=-\alpha W \mathsf{J}$.
The initial conditions at the source are
\begin{equation}
\mathsf{W}=1, \ \mathsf{M}=1, \ \mathsf{J}= \ \! \eta^{-1}. 
\label{ci2}
\end{equation}
From the equations (\ref{ecg}),(\ref{ci2}) it follows that also in the
modified Gaussian model the fountain regime is determined by the
parameters $\eta$ and $\beta$.

We solved the Eqs.  (\ref{ec2}) with the initial conditions
(\ref{ci2}) to obtain the critical conditions for the collapse of the
fountain with Gaussian profile. The resulting critical line is shown
in Fig.\ref{figsigmacgauss}.

\begin{figure}
    \caption{Critical line for the collapse of the fountain in the Gaussian model described by Eqs. (\ref{ecg}), (\ref{nc}), with $\gamma=0$ (solid line) and $\gamma=0.25$ (dotted line).}
    \label{figsigmacgauss}
\end{figure}

\section{Top-Hat-Gaussian model}
Although the \textit{MTT} equations have been formulated for the cases
of top-hat and Gaussian profiles, in real situations it is observed
that in many cases the profiles are not purely top-hat nor Gaussian
\cite{Lee}. This fact motivates us to consider a model in which the
velocity of the jet is a combination of these two types of flows. Then
we assume that the velocity and density are
\begin{eqnarray}
\tilde{u}(z,r) &=&(A e^{-r^2 / b^2} + B (1-\Theta(r-b))) u(z) \label{uc2}, \nonumber\\
\tilde{\rho}(z,r)&=&A((\rho(z)-\rho_0) e^{-r^2 / b^2}+\rho_0) \nonumber \\ &+& B (\rho + (\rho_0-\rho)\Theta(r-b)). 
\label{ecm1}
\end{eqnarray}
where $\Theta(x)$ is the Heaviside function, defined as $\Theta(x)=0$
if $r<b$, $\Theta(x)=1$ if $r \ge b$, and $u$ is the characteristic
velocity, which corresponds to the velocity of the jet at the
axis. Thus, in the model $A+B=1$. To characterize the flow we
introduce the parameter $\delta=1-A$. Hence, if $\delta=0$ the model
reduces to the Gaussian profile and if $\delta=1$ the top hat model is
recovered. As done in \cite{Morton1956}, we integrated the equation of
momentum conservation in an infinite section to eliminate the
dependence with $r$. Assuming again the Boussinesq condition, it is
obtained that
\begin{equation}
C_1 \frac{d}{dz}(\pi \rho b^2 u^2)= -g b^2 (\rho-\rho_0)    
\end{equation}
where $C_1=0.5 A^3+1.764 A^2 B + 3 AB^2 + B^3$. Thus the resulting equation is similar to the original \textit{MTT} equations, except for the factor $C$. Then, it follows that this equation can be written as
\begin{equation}
{dM^2 \over dz}=2 C^{-1} F W. 
\end{equation}
On the other hand, we obtained an equation similar to (\ref{ed3}) for
the mixed flow,
\begin{equation}
\frac{dG}{dz}=-  D \alpha W^{-1}M^{1/2}G+N^2,
\end{equation}
 where $C_2=(0.264A+B)/(0.316A+0.5 B)$. Since the accepted values of
 $\alpha$ are near $0.06$ and $0.08$ for the Gaussian and top-hat
 profiles respectively, we assume that in the model
 $\alpha=0.06+0.02\delta$. This assures that the usual values are
 recovered in the limiting cases $\delta=0$ (Gaussian) and $\delta=1$
 (top hat). Alternatively, the value of $\alpha$ could be considered
 to be a free parameter. Thus, the dimensionless equations of the
 model are
\begin{eqnarray}
{ d\mathsf{W} \over d\mathsf{x}} &=&2 \mathsf{M}^{1/2}, \nonumber \\  
  {d\mathsf{M}^2 \over d\mathsf{x}}&=&-2 C_1^{-1} \mathsf{J} \mathsf{W}^2,  \nonumber \\
 {d\mathsf{J} \over d\mathsf{x}}=&-& C_2 {\mathsf{W}}^{-1}\mathsf{M}^{1/2}\mathsf{J}+\beta^2 ,
\label{ecm}
\end{eqnarray}
with the initial conditions 
\begin{equation}
\mathsf{W}=1, \ \mathsf{M}=1, \ \mathsf{J}= \ \! \eta^{-1}. 
\label{ci3}
\end{equation}

\subsection{Numerical simulations}
\begin{figure}
    \includegraphics[width=90mm]{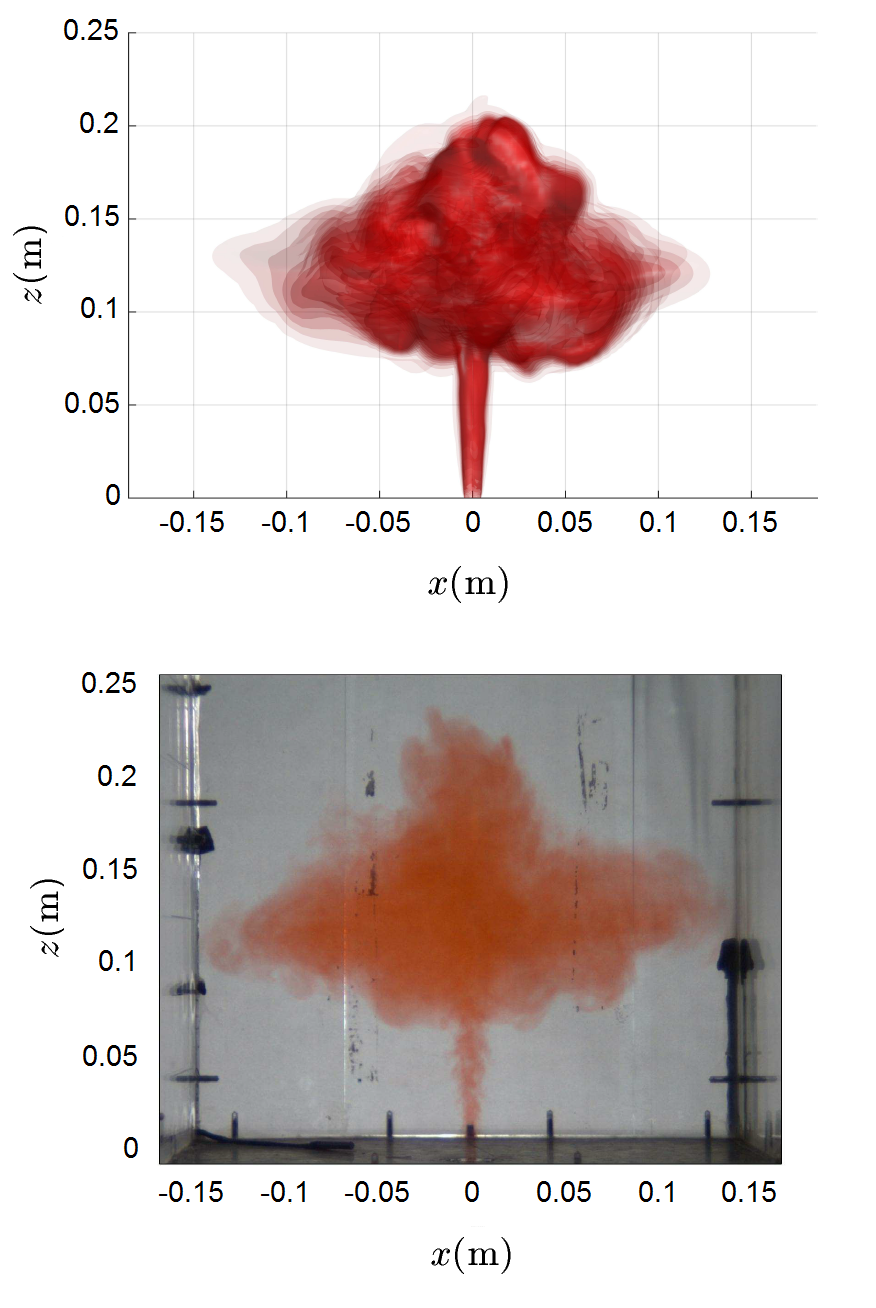}
    \caption{Comparison between the experiments and simulations of the fountain spreading in the zero buoyancy case. The buoyancy frequency is $N=0.23$ s$^{-1}$, the radius of the nozzle $d=4$ mm and the velocity at the inlet $U=0.11$ m/s.}
    \label{figsim_t}
\end{figure}

\begin{figure}
    \includegraphics[width=90mm]{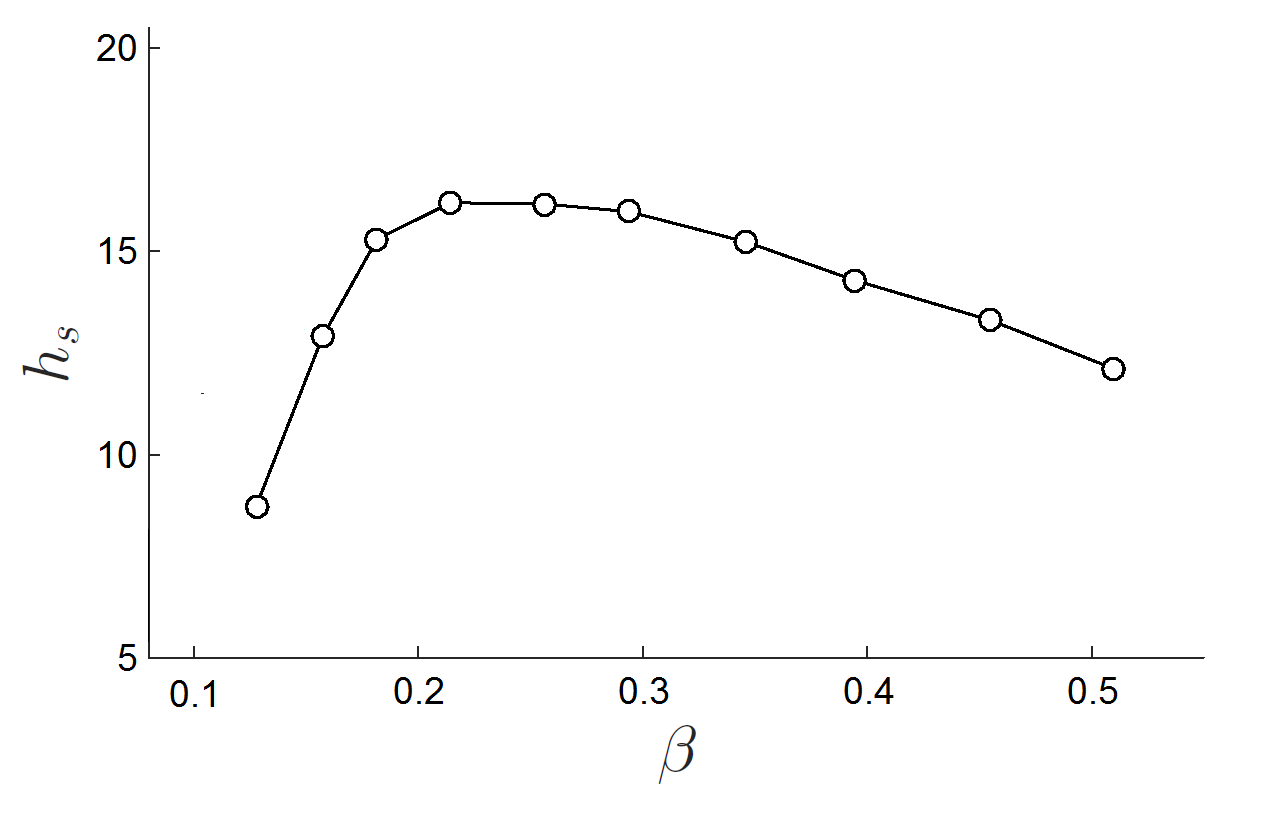}
    \caption{Dimensionless height $h_s$ as a function of $\beta$ for $\eta=8$ obtained with the numerical simulations in the nonzero buoyancy case. As it can be seen, $h_s$ exhibits a maximum when it is plotted as a function of $\beta$ at constant $\eta$, as it is predicted by the model.}
    \label{fighssim}
\end{figure}

Numerical simulations were implemented to obtain an estimation of the
value of the parameter $\gamma$ that has been introduced to take into
account the mixing in the downflow and to test other predictions of
the model.  The simulations were performed with the program
caffa3d.MBRi, which is an open-source code that implements the Finite
Volume Method (FVM).  This solver has been tested for accuracy in
benchmark flows and it has been shown to have second-order accuracy in
space and time.  The code solves numerically three-dimensional
incompressible flows using curvilinear meshes structured by
blocks. For more information about the solver we refer to
\cite{Usera},\cite{Mendina}. To validate the computational scheme, we
contrasted the results with the experimental data of Freire et. al
\cite{Freire2010}. These experiments were conducted in a prismatic
container with lateral walls of 0.4 m in width and 1 m in height. The
ambient fluid was water and the density stratification was established
fixing a temperature of $40^{\circ}$C at the top and $15^{\circ}$C at
the bottom. The fluid jet was injected upwards through a nozzle of $8$
mm diameter. Differing degrees of turbulence level were generated in
the experiments using a stainless-steel wire mesh placed at the inlet
port.  In the numerical implementation, the generation of disturbances
produced by the wire mesh was simulated by the inclusion of white
noise in the velocity components of the jet at the inlet. The
intensity of the noise was varied to obtain the best fit of
experimental observations.  A comparison of the numerical and
experimental results is given in Fig. \ref{figsim_t}, showing a very
good agreement. These simulations were done using a noise level of
20\% of the velocity magnitude at the inlet. After the validation of
the program, we performed simulations for the non-zero buoyancy flux
case considering different values of the buoyancy frequency $N$ in the
range $0.1 - 0.4$ s$^{-1}$. The inlet velocity $U$ was kept fixed at
$0.11$ m/s and the jet temperature at the inlet was fixed at the value
$T_J=10^{\circ}C$. Using the results of the temperature field and
Eq.(\ref{ecgamma}) the values of $\gamma$ of different runs were
calculated. The average value of the obtained values is $\gamma=0.011
\pm 0.003$.  We also contrasted the model with the results of the
simulation in another way. As it can be observed in
Fig. \ref{figsigmachs}, $h_s$ experiments a maximum when $\beta$ is
varied at a fixed value of $\eta$. The results of the simulation
concerning the value of $h_s$ are plotted in Fig. \ref{fighssim},
showing that the prediction of the model is confirmed by the numerical
simulation.

\subsection{Comparison with experimental data}

In order to compare the results of the different models with the experimental measurements, we first consider the case of zero buoyancy (Fr$\rightarrow \infty$).
 For the sake of convenience, the maximal and spreading heights $h_m$, $h_s$ will be normalized with the characteristic length
$$
l_j= (\pi M_i)^{1/4 }N^{-1/2}
$$
used by Papanicolau et al. \cite{Papanicolau} to plot their experimental data and those reported by  Bloomfield and Kerr \cite{Bloomfield1999}. 
We will  also consider the ratios $z_s/z_m$. In their experimental study, Papanicolau et. al concluded that in the limit $\sigma \rightarrow \infty$, which is the zero-buoyancy limit, the ratio $z_m/l_j$ tends to the same value $z_m/l_j \approx 3.6$. Bloomfield and Kerr reported a similar behavior but with a different value in the zero-buoyancy limit $z_m/l_j \approx 2.88$. Papanicolau et. al attributed these differences to probable errors in the experimental procedure of Bloomfield and Kerr. However, the results of the present models suggest that the explanation can be another. Since $h_m$, $h_s$ do not obey to universal functions of $\sigma$ as it was incorrectly assumed in \cite{Bloomfield1998}, there is not a universal value of  $z_m/l_j$. This value depends on other conditions under which the experiment is done, such as the value of $\beta$. In Table \ref{table1},  the values of $z_m/l_j$, $h_s/h_m$ obtained with the models and in the experiments  are shown.

\begin{table}[h]
\begin{center}
\caption{Experimental data and results of the models for $\mathsf{N}=0.01$.}
\label{table1}
\begin{tabular}{|c|c|c|c|}
\hline
Study/model &  $z_m/l_j$ & $z_s/l_j$  & $z_s/z_m$  \\ \hline
Bloomfield and Kerr       & 2.88          & 1.35 & 0.47                \\ \hline
Papanicolau et al.        & 3.58          & 1.94 & 0.54                \\ \hline

Top-hat, $\gamma=0.11$         &     2.25      &  0.93  &    0.41           \\ \hline

Gaussian, $\gamma=0.5$         & 1.96           & 0.89 &   0.45              \\ \hline
Top-Hat-Gaussian,  $\gamma=0.18$ , $\delta=0.9$         &        2.26   & 0.98 & 0.43                 \\ \hline
\end{tabular}
\end{center}
\end{table}

The results shown in the table were obtained for $\mathsf{N}=0.01$, which is a value that is representative of the conditions of the experiments in \cite{Bloomfield1998}. As it can be seen, for the above mentioned condition the results are close to the data in \cite{Bloomfield1998}. In the case of the Gaussian model, to obtain values near the data a value of $\gamma=0.5$ was used, which  is probably not as realistic as it follows from the estimations  that we did using the simulations. The best results were obtained with the mixed model with a flow that  is closer to the top-hat profile than the Gaussian profile ($\delta=0.9$). We note that the values of $z_m/l_j$, $z_s/l_j$ are not universal but depend on $\beta$. The results in the table are closer to the data of \cite{Bloomfield1998} probably because we used a value of $\beta$ that is in accordance with the conditions of these experiments. 

We now consider the comparison between the critical values  $\sigma_c$ obtained in experiments performed with non-zero buoyancy flux at the source and  those obtained with  the model. In the experiments carried out by Bloomfield and Kerr, it was obtained that $\sigma_c \sim 5$ \cite{Bloomfield1999}. In this work the authors performed a series of experiments varying $N$ and Fr, which implies that different values of $\eta$ were covered. We estimate that a representative value of this parameter in such experiments is $\eta \sim 80$. Using the mixed models with $\delta=0.9$, $\gamma=0.18$ and $\eta=80$ we obtained  $\sigma_c=4.6$ and $z_m/lj=1.94$ at the critical point $\sigma=\sigma_c$, which are close to the values $\sigma_c=5$, $z_m/lj\approx2$ obtained in \cite{Bloomfield1999} (see also \cite{Papanicolau}).

\section{Conclusion}\label{Sec:conclusiones}

In this work we considered models to describe the collapse or  rise and spreading of  axisymmetric turbulent fountains taking into account all the relevant parameters of the flow. In the first considered model we combine the  \textit{MTT} equations for the top-hat profile and the condition (\ref{ehs}) to determine in a first approximation the dimensionless maximal and spreading heights $h_m$, $h_s$. In this model, all the fountain regimes can be specified by the two dimensionless parameters $\beta=\mathsf{N} /\alpha$ and $\eta=\alpha Fr^2$. The results showed that the model describes in correct form at a qualitative level the dependence of $h_m$ and $h_s$ with all the parameters of the flow, including the conditions at the source and of the environment. We obtained the critical conditions to the collapse of the fountain, in which case the flow does not become neutrally buoyant at a positive $h_s$,  but falls to the source level. The dependence of $h_s$  on $\sigma$ is very abrupt around the collapse conditions, in accordance with the experimental data. 
 The results showed that the relation $h_m=f(\sigma) M_0^{3/4}F_0^{-1/2}$ that has been proposed in various previous works \cite{Bloomfield1998,Bloomfield1999,Bloomfield2000} is not correct. The origin of the drawback is that this relation does not take into account all the dimensionless parameters determining the flow. This fact explains some important differences concerning the results of distinct experimental investigations \cite{Bloomfield1998},\cite{Papanicolau}.
 
 Although the predictions of the model concerning the dependence of $h_m$ and $h_s$ with the parameters controlling the phenomena are qualitatively correct, the numerical values do not fit  the experimental data accurately. This discrepancy is caused by the fact that the model does not include the mixing that occurs between the falling flow and the environment after the jet attained the maximal height and reversed its direction. To correct this deficiency we modified the original condition used to estimate $h_s$ to include the mixing in the downflow introducing a parameter $\gamma$. This modification increased the agreement between the predictions and the experimental data. The value of $\gamma$ has been estimated by the use of numerical simulations.
We also considered a  model to describe jets with Gaussian profile. 
We showed that the classics  \textit{MTT} equations in the case of the Gaussian profile  predict an incorrect evolution of the fluid density inside the jet, for which reason the spreading height obtained using Eqs.(\ref{mtt}) and the condition (\ref{ehs}) is always negative, which means that in this model the fountain collapses in all the cases.  We derived a new equation to complete the description, which does not exhibit the problem described above. We show that the new set of equations together with Eq. (\ref{ehs}) determine,  from qualitative viewpoint, the dependence  $h_m$ and the $h_s$ with $\sigma$ correctly. In order to increase the quantitative quality of the model, we introduced a third model which assumes that the flow is a combination of a Gaussian and a top-hat jet. With this mixed model, very good agreements were obtained with the experimental data.  
Thus, we presented a model which captures in qualitative and quantitative form the dependence of the rise and spreading on all the fundamental parameters that characterizes the fountain. As a consequence, the mixed model can be used to predict the overall behavior of fountains in a stratified medium. Other predictions of the model were satisfactorily  verified with the numerical simulations.

\ \


The authors would like to thank the Uruguayan institutions Programa de Desarrollo de las Ciencias B\'asicas (PEDECIBA), Agencia Nacional de Investigaci\'on e Innovaci\'on (ANII)
and express their gratitude for the grant F\'{\i}sica Nolineal (ID 722)
Programa Grupos I+D CSIC 2018 (Udelar, Uruguay).




\end{document}